\documentclass[aps,prl,twocolumn,groupedaddress,showpacs,showkeys]{revtex4-1}
\usepackage{graphicx}
\usepackage{dcolumn}
\usepackage{bm}
\usepackage{amssymb,amsmath,latexsym,amsfonts}
\usepackage{float} 
\usepackage{footnote}
\begin{document}


\title{String theory phenomenology and quantum many--body systems}

\author{S. Guti\'errez}
 \email{sergiogs@xanum.uam.mx} \affiliation{Departamento de F\'{\i}sica,
 Universidad Aut\'onoma Metropolitana--Iztapalapa\\
 Apartado Postal 55--534, C.P. 09340, M\'exico, D.F., M\'exico.}

\author{A. Camacho}
 \email{acq@xanum.uam.mx} \affiliation{Departamento de F\'{\i}sica,
 Universidad Aut\'onoma Metropolitana--Iztapalapa\\
 Apartado Postal 55--534, C.P. 09340, M\'exico, D.F., M\'exico.}

\author{H. R\'{\i}os}
 \email{hrh@xanum.uam.mx} \affiliation{Departamento de F\'{\i}sica,
 Universidad Aut\'onoma Metropolitana--Iztapalapa\\
 Apartado Postal 55--534, C.P. 09340, M\'exico, D.F., M\'exico.}


 \date{\today}

\begin{abstract}
The main idea in the present work is the definition of an
experimental proposal for the detection of the number of
extra--compact dimensions contained as a core feature in String
Theory. This goal will be achieved as a consequence of the fact that
the density of states of a bosonic gas does depend upon the number
and geometry of the involved space--like dimensions. In particular
our idea concerns the detection of the discontinuity of the specific
heat at the condensation temperature as a function of the number of
particles present in the gas. It will be shown that the
corresponding function between these two variables defines a segment
of a straight line whose slope depends upon the number of
extra--compact dimensions. Resorting to some experiments in the
detection of the specific heat of a rubidium condensate the
feasibility of this proposal using this kind of atom is also
analyzed.

\end{abstract}

\maketitle

\date{\today}

\section{\bf Introduction}

Nowadays it is widely accepted that modern physics has an unsolved
problem, i.e, the so--called quantum gravity puzzle, namely, there
is no quantum version of gravity \cite{Amelino1}. This issue has
been addressed for more than seven decades \cite{Stachel}. There is
an element in this context that has to be mentioned, namely, the
path in this direction has no guidance, at all, from any kind of
experiments, a fact stemming from the disparity between our current
technology and the order of magnitude of the effects predicted by
the corresponding models \cite{Amelino2}. Of course, this situation
cannot be considered an advantage in this quest.

The number of ideas in the pursue of a quantum theory of gravity
grows from day to day and includes String Theory, Loop Quantum
Gravity, Non--Commutative Spacetimes, etc. \cite{Amelino1}. In the
context of String Theory the predicted new effects lie completely
outside the testable zone for our technology, a situation
consequence of the fact that their order of magnitude is determined
by the Planck scale, an apparently very pessimistic scenario. At
this point it has to be commented that the proposed experiments
involve always a single quantum system, i.e., a single quantum
particle suffers the effects of the new features predicted by the
involved model \cite{Kane}.

The topic of ultra--cold quantum gases is currently one of the
hottest spots in modern physics \cite{Griffin} and the work in this
realm embodies many possibilities, they range from superfluidity,
low--dimensional systems, etc. The question concerning its
possibilities in quantum gravity phenomenology is an issue that has
to be addresses thoroughly and its limitations and ensuing options
have to be understood. Here we deal with one of the possible windows
that ultra--cold systems could offer us in connection with
gravitation. In the present work we address the issue of the number
of extra compact dimensions from the perspective of a many--body
system at ultra--cold temperatures. In other words, we do not seek
the effects upon a single quantum particle of some new theoretical
element contained in String Theory. The main idea is to exploit some
features contained in the physics of ultra--cold systems which
depend strongly upon the number of space--like dimensions. More
precisely, some thermodynamical quantities, for instance, specific
heat, energy, do show a clear dependence upon the number of
spacelike dimensions of the corresponding model \cite{Landsberg}.

The system to be analyzed will be a gas containing $N$ bosonic
particles, which interact pairwise in repulsive way. The
corresponding properties of this model will be studied in the region
of ultra--cold temperatures, i.e., about the corresponding
condensation temperature. An anisotropic harmonic oscillator will be
present, as is usual in Bose--Einstein Condensation experiments
(BEC) \cite{Andrews1}, and the ensuing thermodynamics will be
deduced assuming that $l$ compact dimensions are also present, the
order of magnitude of these dimensions will be considered equal or
smaller than Planck's length $l_P$ \cite{Amelino3}. This element is
one of the core features of String Theory \cite{Barton}. The effects
of the trap upon the compact dimensions can be neglected. Indeed,
for a gas, with atoms of mass $m$ and frequency of the trap given by
$\omega$, the characteristic length of this system is
$\sqrt{\hbar/(m\omega)}$, since, by hypothesis,
$l_P\ll\sqrt{\hbar/(m\omega)}$, then in any of the compact dimensions
the presence of the trap reduces to a constant potential $V_0$. In
other words, concerning the extra part of the geometry, the
particles behave as a particles in a box in which a constant
potential is present, this last comment implies that the density of
states (as function of the energy) of this situation is the same as
in the case in which $V_0=0$ \cite{Schiff}. The energy of this
system will be calculated and the ensuing specific heat will be
deduced. As is known, even for an ideal bosonic gas in a box, the
specific heat shows a discontinuity at the critical temperature,
such that this jump is a function of the number of spacelike
dimensions \cite{Landsberg}. This last trait will be exploited in
the present work. Indeed, we will deduce the discontinuity of our
bosonic gas as function of the number of atoms present in the gas,
assuming $l$ compact dimensions and $s$ non--compact ones. It will
be shown that the functional dependence between discontinuity and
number of particles defines a segment of a straight line whose slope
is a function of $s$ and $l$.

More precisely, our experimental proposal is the following one: Take
a bosonic gas with $N$ particles and measure the specific heat above
and below the condensation temperature and calculate the
corresponding discontinuity. Repeat this procedure for different
values of $N$. The resulting graph will be the segment of a straight
line whose slope will contain information of the number of extra
compact dimensions, of course, also of the non--compact ones. A
realistic proposal shall include the pairwise interaction present in
any dilute bosonic gas \cite{Pethickbook}. In order to have a clear
interpretation of some thermodynamic parameters, for instance, the
chemical potential at or below the condensation temperature, the
analysis will be done assuming that the pairwise interaction, here
codified in the scattering length $a$, can be handled in the context
of the variational approach \cite{Pethickbook}. This condition leads
us to an inequality relating $N$, $a$ and $\sqrt{\hbar/(m\omega)}$,
i.e., $Na<\sqrt{\hbar/(m\omega)}$. The present status in the
experimental part tells us that most of the situations satisfy the
opposite condition ($Na>\sqrt{\hbar/(m\omega)}$), called the
Thomas--Fermi limit \cite{Henk}, nevertheless, there are experiments
carried out within the validity regime of the variational approach
\cite{Pethickbook}, i.e., our mathematical restriction does not
imply an unrealistic proposal.

\section{\bf Extra--dimensions and density of states for a bosonic gas}
As mentioned before our model is defined by $s$ non--compact
dimensions and $l$ compact ones such that the latter have a size
denoted by $R_j$, with $j=1,2,...,l$, and all of them have an order of
magnitude similar to Planck length, i.e., $R_j\sim l_p$; this is one
of the main assumptions of String Theory \cite{Barton}.

In this geometry we consider bosonic particles with mass $m$ trapped
by a harmonic potential that has non--vanishing frequencies (here
denoted by $\omega_i, ~i=1, 2,...,s$) only in the non--compact
space--like dimensions, in other words, the trap has no effects upon
the compact part of the geometry. The energy of a particle in this
situation will be determined by its contribution from the
non--compact dimensions plus the part of the compact ones. Clearly,
we have

\begin{eqnarray}
\epsilon=\hbar[\omega_1(n_1+1/2)+...+\omega_s(n_s+1/2)]\nonumber\\
+\frac{\pi^2\hbar^2}{2m}[\frac{q^2_1}{R^2_1}+...+\frac{q^2_l}{R^2_l}].
\label{equa01}
\end{eqnarray}

In this last expression $n_i$ and $q_j$ denote the quantum numbers
associated to the harmonic oscillator potential and the case of a
particle immersed in a box of size $R_j$, respectively. In other
words, the energy of a particle is the sum of the contributions
stemming from the harmonic trap and that emerging from the fact that
in the extra--dimensions the particle shows the behavior of a free
particle \cite{Barton}.

A main element is the density of states of a single particle
immersed in the aforementioned geometry, with it we may deduce the
thermodynamics of our system \cite{Ueda}. Let us address the issue
of the number of states, here denoted by $G(\epsilon)$, whose energy
is equal or smaller than a certain value $\epsilon$. The answer to
this question is, in the semiclassical approximation
\cite{Pathria1}, proportional to the volume of a figure with two
independent contributions, namely, one stemming from an
hyper--ellipsoid in $l$ dimensions and the usual case for a harmonic
trap in $s$ non--compact dimensions \cite{Pethickbook}. The analysis
of the compact dimensions is equal to the case of a gas immersed in
an $l$--dimensional box, and the deduction of the corresponding
number of states appears in any text on Statistical Mechanics
\cite{Landsberg, Pathria1}.

The integral to be calculated reads

\begin{eqnarray}
G(\epsilon)=\frac{1}{2^l}\int_0^{c_1}dn_1\int_0^{c_2}dn_2...\int_0^{c_s}dn_s\int_0^{\tilde{Q}}S_lr^{l-1}dr.
\label{equa02}
\end{eqnarray}

Here we have that $\tilde{Q}=
\frac{2m\tilde{R}^2}{\pi^2\hbar^2}\epsilon$, the radius (in the
$q$--space) of the hyper--sphere, see below the definition of this
hyper--sphere in terms of the hyper--ellipsoid. The term $1/2^l$
stems from the fact that concerning the compact dimensions only that
part of the geometry (in which all the quantum numbers $q_j$ are
non--negative) is to be taken into account \cite{Landsberg,
Pathria1}. In addition, we have resorted to hyper--spherical
coordinates in which the volume of the hyper--ellipsoid equals that
of a hyper--sphere whose radius is the geometric average of the
semi--axis of the hyper--ellipsoid, namely,

\begin{eqnarray}
\tilde{R}=[R_1R_2...R_l]^{1/l} \label{equa03}
\end{eqnarray}

Furthermore, the term $S_l$ appears as a consequence of the
integration in hyper--spherical coordinates of the angle parameters
\cite{Barton}

\begin{eqnarray}
S_l=\frac{2\pi^{l/2}}{\Gamma(l/2)}.
\label{equa04}
\end{eqnarray}

Here $\Gamma$ denotes Euler gamma function. Clearly, as is known for
the case of an ideal gas in an $l$--dimensional box \cite{Pathria1},
$\tilde{Q}^2$ defines in the space of the quantum numbers $q_j$ the
radius of our hyper--sphere. For our case, the functional dependence
of $\tilde{R}$ with $\epsilon$ and the involved quantum numbers is
provided by

\begin{eqnarray}
q^2_1+...+
q^2_l=\frac{2m\tilde{R}^2}{\pi^2\hbar^2}\Bigl(\epsilon-\hbar[\omega_1n_1+...+\omega_sn_s]\Bigr)
\label{equa05}
\end{eqnarray}

Finally, we have that

\begin{eqnarray}\label{equa06}
c_s=-\frac{1}{\hbar\omega_s}\Bigl(\epsilon-\hbar[\omega_1n_1+...+\omega_{(s-1)}n_{(s-1)}]\Bigr)
\end{eqnarray}

\begin{eqnarray}
c_1=\frac{\epsilon}{\hbar\omega_1}.
\label{equa07}
\end{eqnarray}

With these previous remarks we may now calculate (\ref{equa02})
(here $\tilde{\omega}$ denotes the geometric average of the
frequencies)

\begin{eqnarray}
G(\epsilon)
=\frac{\pi^{l/2}}{2^l\Gamma(l/2+s+1)}\Bigl(\frac{2m\epsilon\tilde{R}^2}{\pi^2\hbar^2}\Bigr)^{l/2}\Bigl(\frac{\epsilon}{\hbar\tilde{\omega}}
\Bigr)^s.\label{equa08}
\end{eqnarray}

The density of states is provided by
$\Omega(\epsilon)=dG(\epsilon)/d\epsilon$

\begin{eqnarray}
\Omega(\epsilon)
=\frac{\pi^{l/2}}{2^l\Gamma(l/2+s)}\Bigl(\frac{2m\epsilon\tilde{R}^2}{\pi^2\hbar^2}\Bigr)^{l/2}\frac{\epsilon^{s-1}}{(\hbar\tilde{\omega})^s}
.\label{equa09}
\end{eqnarray}

At this point let us consider some limit cases. For instance, if we
fix $l=0$ and $s=3$ our last expression reduces to

\begin{eqnarray}
\Omega(\epsilon)
=\frac{1}{2}\frac{\epsilon^{2}}{(\hbar\tilde{\omega})^3}
.\label{equa10}
\end{eqnarray}

Clearly, we recover an already known result \cite{Stringaribook1},
the density of states of a bosonic gas trapped by an anisotropic
harmonic oscillator. On the other hand, if we consider the case of
gas trapped in a three--dimensional box, we end up with the case
which, in connection with the density of states, is equal to the
case of a gas in a three--dimensional compact geometry. In other
words, the conditions $l=3$  and $s=0$ shall imply the deduction of
a gas as trapped in a three--dimensional box. Expression
(\ref{equa09}) leads us to confirm our conjecture \cite{Pathria1}

\begin{eqnarray}
\Omega(\epsilon)
=\frac{\pi^{3/2}}{\Gamma(3/2)}\Bigl(\frac{m\epsilon\tilde{R}^2}{2\pi^2\hbar^2}\Bigr)^{3/2}
.\label{equa11}
\end{eqnarray}

Our result also states that if we start with a three--dimensional
BEC (assuming the absence of non--compact dimensions) then the
density of states is provided by (\ref{equa10}). If we now restrict
the motion of the particles along any of the available axis, then
the resulting density of states is provided by setting $s=2$ and
$l=1$, it is not the case of two non--compact dimensions
\cite{Mullin} in which $s=2$ and $l=0$. Of course, (\ref{equa09})
contains the purely two--dimensional situation, we must fix $s=2$
and $l=0$. In other words, restricting the motion along a
non--compact dimension is not equivalent to the case of a BEC in
which the universe has one less dimension. The existence of
restricted motion along any axis entails that the energy eigenvalues
are the same as those stemming from a compact dimension with the
same size, of course, they impinge upon the density of states,
unavoidably.

 The number of excited states available to the gas is given by
($z=\exp{(\mu\beta)}$, here $\mu$ is the chemical potential and
$\beta=1/(\kappa T)$, with $\kappa$ Boltzmann's constant
\cite{Pathria1})

\begin{eqnarray}
N_e=\int_0^{\infty}\frac{\Omega(\epsilon)}{z^{-1}\exp{(\beta\epsilon)}-1}d\epsilon
.\label{equa12}
\end{eqnarray}

The energy reads

\begin{eqnarray}
E=\int_0^{\infty}\frac{\epsilon
\Omega(\epsilon)}{z^{-1}\exp{(\beta\epsilon)}-1}d\epsilon
.\label{equa13}
\end{eqnarray}

 We may now proceed to calculate the condensation temperature and the energy
of the system. As usual, the critical temperature is deduced
imposing the condition that the number of particles equals $N_e$ and
that the chemical potential at the critical temperature ($\mu_c$)
equals the smallest of the energy eigenvalues of our system
\cite{Stringaribook1}. In the general case there is no known answer
to the question concerning the eigenvalues of those particles in a
BEC \cite{Lieb}, but in the variational approach the presence of a
non--vanishing pairwise interaction entails only a re-scaling of the
frequencies of the trap \cite{Pethickbook}, the order parameter,
after the introduction of a non--vanishing scattering length,
continues to have the structure determined by a harmonic oscillator,
but now the presence of a repulsive interaction ($a>0$) entails that
it becomes larger. Under this assumption, we have that the chemical
potential at the critical temperature, or below it, reads
$\mu_c=(\hbar/2)[\hat{\omega}_x+\hat{\omega}_y+\hat{\omega}_z]$),
 where now these effective frequencies are provided by
 ($\tilde{L}=(L_xL_yL_z)^{1/3}$, $\tilde{\omega}=(\omega_x\omega_y\omega_z)^{1/3}$, with $L_j=\sqrt{\frac{\hbar}{m\omega_j}}$)

\begin{eqnarray}
{\sqrt\frac{\hbar}{m\hat{\omega}_j}}=
\Bigl(\frac{2}{\pi}\Bigr)^{1/10}\Bigl(\frac{Na}{\tilde{L}}\Bigr)^{1/5}\frac{\tilde{\omega}}{\omega_j}\tilde{L}.\label{equa14}
\end{eqnarray}

The variational approach allows us have an analytical result for
$\mu_c$. This last expression is valid only if $Na<L_j$.

We may now deduce the condensation temperature, here $g_{\nu}(x)$
are the so--called Bose functions \cite{Pathria1}

\begin{eqnarray}
N=\frac{\pi^{l/2}}{2^l}\Bigl(\frac{2m\kappa
T_c\tilde{R}^2}{\pi^2\hbar^2}\Bigr)^{l/2}\Bigl(\frac{\kappa
T_c}{\hbar\tilde{\omega}}\Bigr)^sg_{\nu}(x).\label{equa15}
\end{eqnarray}

In this last expression we have that, concerning the Bose function,
$\nu=s+l/2$ and $x=\frac{\mu_c}{\kappa T_c}$. In order to have an
analytical expression for the critical temperature let us recall
that our expressions (starting with the deduction of the density of
states) are valid in the semi--classical limit
($(1/2)\hbar[\omega_x+\omega_y+\omega_z]<\kappa T_c$). This last
comment allows us to approximate our Bose function
($g_{\nu}(x)=\sum_{l=1}^{\infty}\frac{x^l}{l^{\nu}}$) as follows
($\zeta(x)$ denotes Riemann's zeta--function \cite{Pathria1})

\begin{eqnarray}
N=\frac{\pi^{l/2}}{2^l}\Bigl(\frac{2m\kappa
T_c\tilde{R}^2}{\pi^2\hbar^2}\Bigr)^{l/2}\zeta(s+l/2)
\Bigl(\frac{\kappa
T_c}{\hbar\tilde{\omega}}\Bigr)^s\Bigl(1\nonumber\\
+\frac{\zeta(s-1+l/2)}{\zeta(s+l/2)}\frac{\mu_c}{\kappa
T_c}).\label{equa16}
\end{eqnarray}

We now proceed to calculate the energy of the system. Below the
critical temperature the chemical potential is a constant
\cite{Stringaribook1}, therefore, we deduce the energy in two parts.

Firstly, we take $T>T_c$, then the corresponding integration renders

\begin{eqnarray}
E=(l/2+s)\frac{\pi^{l/2}}{2^l}\Bigl(\frac{2m\kappa
T\tilde{R}^2}{\pi^2\hbar^2}\Bigr)^{l/2}g_{(s+1+l/2)}(z)\nonumber\\
\Bigl(\frac{\kappa
T}{\hbar\tilde{\omega}}\Bigr)^s\kappa T.\label{equa17}
\end{eqnarray}

Secondly, for $T<T_c$, we conclude

\begin{eqnarray}
E=(l/2+s)\frac{\pi^{l/2}}{2^l}\Bigl(\frac{2m\kappa
T\tilde{R}^2}{\pi^2\hbar^2}\Bigr)^{l/2}\nonumber\\
\zeta(s+1+l/2)(z)\Bigl(\frac{\kappa
T}{\hbar\tilde{\omega}}\Bigr)^s\kappa T\nonumber\\
\Bigl(1 +\frac{\zeta(s+l/2)}{\zeta(s+1+l/2)}\frac{\mu_c}{\kappa
T_c}).\label{equa18}
\end{eqnarray}

A fleeting glimpse at these last two expressions allows us to notice
that they depend upon the number of non--compact dimensions, but
also on the number of extra dimensions. This fact is the main
ingredient in our proposal.

In the context of BEC it is known that the specific heat shows a
discontinuity at the critical temperature \cite{Stringaribook1}, a
fact present even in the liquefaction of helium, and referred as a
$\lambda$ transition \cite{Nozieres}.

These last two expressions allow us to deduce the specific heats
(one above the critical temperature and, the second one, below it)
at constant $N$ and constant $\tilde{\omega}$,
$C_{\tilde{\omega}}=(\frac{\partial E}{\partial
T})_{(\tilde{\omega},N)}$.

It will be shown that, for our model, this discontinuity bears
information concerning the number of space--like dimensions, either
non--compact or compact. Indeed,

\begin{eqnarray}
C_{\tilde{\omega}}=\frac{E}{T}\Bigl((s+1+l/2)\nonumber\\
-(s+l/2))\frac{(g_{(s+l/2)}(z))^2}{g_{(s+l/2-1)}(z)g_{(s+l/2+1)}(z)}\Bigr),
~~T>T_c.\label{equa19}
\end{eqnarray}

On the other hand

\begin{eqnarray}
C_{\tilde{\omega}}=\frac{\pi^{l/2}}{2^l}\Bigl(\frac{2m\kappa
T\tilde{R}^2}{\pi^2\hbar^2}\Bigr)^{l/2}\Bigl(\frac{\kappa
T}{\hbar\tilde{\omega}}\Bigr)^s\nonumber\\
\Bigl(l/2+s\Bigr)\Bigl(l/2+s+1\Bigr)\zeta(l/2+s+1)\kappa\nonumber\\
\Bigl[1+ \frac{\zeta(s+l/2)}{\zeta(s+l/2+1)}\frac{\mu_c}{\kappa
T}\Bigr], ~~~T<T_c.\label{equa20}
\end{eqnarray}

The discontinuity is defined as

\begin{eqnarray}
\Delta C_{\tilde{\omega}}= \lim_{T\rightarrow
T_{c+}}C_{\tilde{\omega}}- \lim_{T\rightarrow
T_{c-}}C_{\tilde{\omega}}\label{equa21}
\end{eqnarray}

In order to simplify our expression we resort to (\ref{equa15}) and
find

\begin{eqnarray}
\Delta C_{\tilde{\omega}}=
-N\kappa\Bigl(s+l/2\Bigr)^2\Bigl[\frac{\zeta(s+l/2)}{\zeta(s+l/2-1)}
\nonumber\\
-\frac{\mu_c}{(s+l/2)\kappa T_c}\Bigr].\label{equa22}
\end{eqnarray}

\section{\bf Discussion}

This last expression is our main result and it shows an explicit
dependence upon the number of non--compact dimensions. For the
particular case $s=3$ and $l=0$ we recover the situation for the BEC
in an anisotropic harmonic trap \cite{Stringaribook1}.

\begin{eqnarray}
\Delta C_{\tilde{\omega}}=
-9N\kappa\Bigl[\frac{\zeta(3)}{\zeta(2)}-\frac{\mu_c}{3\kappa
T_c}\Bigr].\label{equa23}
\end{eqnarray}

Furthermore, imposing $s=0$ and $l=3$ we recover the case of a gas
in a three-dimensional box, in which (for this case $\mu_c=0$) no
discontinuity exists \cite{Pathria1}.

\begin{eqnarray}
\Delta C_{\tilde{\omega}}= 0.\label{equa24}
\end{eqnarray}

If we set $s=3$ and $l=6$  (the case of smallest number of
extra--dimensions allowed by String Theory \cite{Barton}), then

\begin{eqnarray}
\Delta C_{\tilde{\omega}}= -36N\kappa\Bigl[0.98-\frac{\mu_c}{6\kappa
T_c}\Bigr].\label{equa25}
\end{eqnarray}

The experimental realm tell us that \cite{Pethickbook} (assuming
$N\sim 10^3$)

\begin{eqnarray}
\frac{\mu_c}{\kappa T_c}\sim N^{-1/3}\Rightarrow
\frac{\mu_c}{6\kappa T_c}\sim 10^{-2}.\label{equa26}
\end{eqnarray}

These last remarks entail that $\Delta C_{\tilde{\omega}}$ is (in
the roughest approximation) a linear function of $N$, and the error
in this model is of 1 percent. The magnitude of the error just
derived matches the experimental uncertainty mentioned in the
literature in connection with the detection of some properties of a
BEC of rubidium \cite{Ensher}.

The experimental proposal is the following one. Measure the
discontinuity in the specific heat of the condensate as a function
of the number of particles. The graph has to be the segment of a
straight line, whose slope shall contain information about the
number of non--compact dimensions. Any deviation from the slope of
$b=-9\zeta(3)/\zeta(2)$ has to be considered as information hinting
to the possibility of $l\not=0$.

The feasibility of the present idea depends upon several
experimental aspects, one of them concerns the fact that the
uncertainty in the detection of the specific heat (here denoted by
$\delta C_{\tilde{\omega}}$) has to be smaller than $\Delta
C_{\tilde{\omega}}$, otherwise the sought effect would be hidden
within the experimental error.

In order to estimate the feasibility we require an experiment in
which not only the aforementioned discontinuity is measured but, in
addition, the corresponding experimental uncertainty is provided.
The only case, known to the authors, fulfilling these requirements
is related to helium \cite{Hull}. The corresponding uncertainty in
the detection of the specific heat depends upon the quantity of mass
involved and reads

\begin{eqnarray}
\delta
C_{\tilde{\omega}}=12.3\times10^{-4}\mathrm{J}/(\mathrm{gm}
^{\circ}\mathrm{K}).\label{equa27}
\end{eqnarray}

We take a BEC of rubidium, namely,  $a\sim 10^{-9}\mathrm{m}$ and
$\sqrt{\frac{\hbar}{m\tilde{\omega}}}\sim  10^{-5}\mathrm{m}$
\cite{Myatt, Chapovsky}. Additionally, our approach requires the
fulfillment of the condition
$Na<\sqrt{\frac{\hbar}{m\tilde{\omega}}}$, notice that $N\sim 10^3$
is a viable value, hence the uncertainty related to this situation
is

\begin{eqnarray}
\delta C_{\tilde{\omega}}=2.75\times 10^{-3}\kappa.\label{equa28}
\end{eqnarray}

For $s=3$ and $l=6$, and the previous values for rubidium, we have

\begin{eqnarray}
\Delta C_{\tilde{\omega}}=-3.5\times 10^{4}\kappa.\label{equa29}
\end{eqnarray}

Clearly, $\vert\Delta C_{\tilde{\omega}}\vert\gg\delta
C_{\tilde{\omega}}$. Since the experimental uncertainty is much
smaller than the effect we conclude that we have a feasible
proposal. There is a work in which the specific heat of a rubidium
gas has been detected \cite{Ensher}, of course, it shows the
mentioned discontinuity. Nevertheless, it is noteworthy to add that
the value of the uncertainty related to the detection in this
parameter is not reported; moreover, the experiment was carried out
for just one value of $N$.

The present model considers the presence of a non--vanishing
pairwise interaction, codified in the scattering length $a\not=0$,
in the context of the validity regime of a variational calculation
\cite{Pethickbook}. This condition implies the fulfillment of

\begin{eqnarray}
Na<\sqrt{\frac{\hbar}{m\tilde{\omega}}}.\label{equa30}
\end{eqnarray}

This imposes an upper bound for the number of particles that can be
used without violating the requirements of the variational approach.
Since our proposal defines $N$ as the independent variable we seek
the largest possible interval for it. This can be done resorting to
Feshbach resonances \cite{Cornish}. Indeed, the interaction in a BEC
can be manipulated when the total energy of a pair of colliding
atoms equals the energy of a quasi--bond state of a molecule, a fact
that leads to the resonant formation of the latter case. In other
words, it implies a change in the corresponding scattering length.
In the very particular case of rubidium \cite{Donley} this option
opens up the possibility of going from $N=80$ up to $N=10^{4}$.
Clearly, the restriction imposed upon $N$ by the mathematical
conditions defining the perturbative approach (see (\ref{equa30}))
has to be fulfilled, nevertheless, we may have a larger interval for
$N$ with the use of Feshbach resonances.

Summing up, we have put forward an experimental proposal, using
bosonic particles, which allows us to determine if our universe
includes extra non--compact dimensions. This has been obtained
resorting to the quantum effects of a many body system such that the
discontinuity of the specific heat has to be measured as a function
of the number of particles and the sought information will be
encoded in the slope of the corresponding graph.

It has to be stressed the difference between the present proposal
and the usual phenomenology in found in the literature of String
Theory \cite{Kane}, namely, the current experimental proposals focus
on a single quantum particle and the effects upon it of the new
physics due to the corresponding model. The difference here is that
we address the issue of a many--body quantum system and exploit the
collective effects that depend upon the number of spacelike
dimensions. In this sense, the present model advocates the topic of
phenomenology of String Theory resorting to quantum system
containing many particles, such as ultra--cold bosonic gases.

\begin{acknowledgements}
H. R. acknowledges CONACyT grant No. 596978 and S. G. the received
UAM grant.
\end{acknowledgements}


\begin{thebibliography}{}
\bibitem{Amelino1}
G. Amelino--Camelia, Living Rev. Relativity \textbf{16}, 5 (2013):
http://www.livingreviews.org/lrr-2013-5.

\bibitem{Stachel}
 J. Stachel, in \emph{Black Holes, Gravitational Radiation and the Universe: Essays in Honour of C. V. Vishveshwara}, B. R. Iyer and B. Bhawal, eds.,
 (Kluwer, Boston) 1999.

\bibitem{Amelino2}
G. Amelino--Camelia, in \emph{Towards Quantum Gravity, Proceeding of
the XXXV International Winter School on Theoretical Physics}, G.
Kowalski--Glikman, ed., (Springer--Verlag, New York) 2000.

\bibitem{Kane}
G. Kane, in \emph{Perspectives on String Phenomenology}, B. Acharya,
et al, eds., (World Scientific Publishing Co., Singapore) 2015.

\bibitem{Griffin}
A. Griffin, T. Nikuni, and E. Zaremba, \emph{Bose--Condensed Gases
at Finite Temperatures}, (Cambridge University Press, Cambridge)
2009.

\bibitem{Landsberg}
 P. T. Landsberg, \emph{Thermodynamics and Statistical Mechanics}, (Dover Publications Inc., New York) 1965.

\bibitem{Andrews1}
M. Andrews, et al, Phys. Rev. Lett. {\bf 79}, 553 (1997).

\bibitem{Amelino3} G. Amelino-Camelia, et al, Int. J. Mod. Phys. D {\bf 19}, 2385 (2010).

\bibitem{Barton} B. Zwiebach, \emph{A First Course in String Theory}, (Cambridge University Press, Cambridge) 2009.

\bibitem{Schiff} L.I. Schiff, \emph{Quantum Mechanics}, (Pergamon Press, Oxford) 1973.

\bibitem{Pethickbook}
C. J. Pethick and H. Smith, \emph{Bose--Einstein Condensation in
Dilute Gases}, (Cambridge University Press, Cambridge) 2008.

\bibitem{Henk}
T.C. Henk, et al, \emph{Ultra--Cold Quantum Fields},
(Springer--Verlag, Dordrecht) 2009.

\bibitem{Ueda}
 M.~Ueda, \emph{Fundamentals and New Frontiers of Bose--Einstein Condensation}, (World Scientic,
 Singapore) 2010.

\bibitem{Pathria1}
 R. K. Pathria, \emph{Statistical Mechanics}, (Butterworth--Heinemann, Oxford) 1996.

\bibitem{Stringaribook1}
 L.~Pitaevskii and S.~Stringari, \emph{Bose--Einstein Condensation}, (Oxford Science Publications,
 Oxford) 2003.

\bibitem{Mullin}
W. J. Mullin, J. Low Temp. Phys.  \textbf{110}, 167--172 (1998).

\bibitem{Lieb}
E. H. Lieb, et al, \emph{The Mathematics of the Bose Gas and its
Condensation}, (Birkh¨\"auser--Verlag, Berlin) 2005.

\bibitem{Nozieres}
P. Nozi\'eres and D. Pine, \emph{The Theory of Quantum Liquids},
Vol. II. (Westview Press, Boulder) 1989.

\bibitem{Ensher}
J. R. Ensher, et al, Phys. Rev. Lett, \textbf{77}, 4984--4987
(1996).

\bibitem{Hull}
R. A. Hull, et al, Proc. Phys. Soc. A \textbf{64}, 379--388 (1951).

\bibitem{Myatt}
Ch. J. Myatt, \emph{Bose--Einstein Condensation Experiments in a
Dilute Vapor of Rubidium}, PhD. Thesis, University of Colorado
(1997).

\bibitem{Chapovsky}
P. L. Chapovsky, JETP Letters. \textbf{95}, 132--136 (2012).

\bibitem{Cornish}
S. L. Cornish and D. Cassettari, Phil. Transac., \textbf{361},
2699--2713 (2003).

\bibitem{Donley}
E. A. Donley, et al, Nature, \textbf{412}, 295--299 (2001).

\end{thebibliography}
\end{document}